\documentstyle[12pt,aaspp4,psfig]{article}
\def\ltsima{$\; \buildrel < \over \sim \;$}
\def\simlt{\lower.5ex\hbox{\ltsima}}
\def\gtsima{$\; \buildrel > \over \sim \;$}
\def\simgt{\lower.5ex\hbox{\gtsima}}
\def\msun{{\rm\,M_\odot}}
\def\lsun{{\rm\,L_\odot}}

\def\s{\ifmmode \widetilde \else \~\fi}
\def\={\overline}

\def\spose#1{\hbox to 0pt{#1\hss}}

\def\etal{{\it et al.\ }}

\def\lta{\mathrel{\spose{\lower 3pt\hbox{$\mathchar"218$}}
     \raise 2.0pt\hbox{$\mathchar"13C$}}}
\def\gta{\mathrel{\spose{\lower 3pt\hbox{$\mathchar"218$}}
     \raise 2.0pt\hbox{$\mathchar"13E$}}}
\def\Dt{\spose{\raise 1.5ex\hbox{\hskip3pt$\mathchar"201$}}}    % upper case
\def\dt{\spose{\raise 1.0ex\hbox{\hskip2pt$\mathchar"201$}}}    % lower case

\def\=={\equiv}

\def\dotsfill{\leaders\hbox to 1em{\hss.\hss}\hfill}

\newcommand{\ffffff}[1]{\mbox{$#1$}}
\newcommand{\scnd}{\mbox{\ffffff{''}\hskip-0.3em .}}
\newcommand{\scmd}{\mbox{\ffffff{''}}}

\newcommand{\name}{APM 08279+5255}

\newcommand{\four}{\rm 450\mu m}
\newcommand{\eight}{\rm 850\mu m}
\newcommand{\thirteen}{\rm 1350\mu m}
\newcommand{\K}{${\rm ^oK}$}

\slugcomment{Submitted to the Astrophysical Journal Letters}

\lefthead{Lewis et al.}
\righthead{\name}

\begin{document}

\title{Submillimeter Observations of the Ultraluminous BAL Quasar \name}

\author{Geraint F. Lewis\altaffilmark{1}
Scott C. Chapman\altaffilmark{2},
Rodrigo A. Ibata\altaffilmark{3},
Michael J. Irwin\altaffilmark{4} \nl \&
Edward J. Totten\altaffilmark{5}}

\altaffiltext{1}{ 
Fellow of the Pacific Institute for Mathematical Sciences 1998-1999, \nl
Dept. of Physics and Astronomy, University of Victoria, Victoria, B.C., Canada 
\& \nl 
Astronomy Dept., University of Washington, Seattle WA, U.S.A.
\nl
Electronic mail: {\tt gfl@uvastro.phys.uvic.ca} \nl 
Electronic mail: {\tt gfl@astro.washington.edu}}

\altaffiltext{2}{Department of Physics and Astronomy, 
University of British Columbia, Vancouver, B.C., Canada \nl
Electronic mail: {\tt schapman@geop.ubc.ca}}
 
\altaffiltext{3}
{European Southern Observatory, Garching bei M\"unchen, Germany \nl
Electronic mail: {\tt ribata@eso.org}}

\altaffiltext{4}
{Royal Greenwich Observatory,
Madingley Rd, Cambridge, UK \nl
Electronic mail: {\tt mike@ast.cam.ac.uk}}
 
\altaffiltext{5}
{Department of Physics, Keele University, Keele,
Staffordshire, UK \nl
Electronic mail: {\tt ejt@astro.keele.ac.uk}}
 
\begin{abstract}
With an inferred bolometric luminosity of $5\times10^{15}{\rm \lsun}$,
the recently identified z=3.87, broad absorption line quasar \name\ is
apparently the most luminous object  currently known. As half of  its
prodigious emission occurs in the infrared, \name\ also represents the
most  extreme  example of an  Ultraluminous  Infrared Galaxy. Here, we
present  new submillimeter  observations  of  this phenomenal  object;
while indicating  that a vast quantity of  dust is present, these data
prove  to be incompatible  with current models  of emission mechanisms
and  reprocessing    in ultraluminous   systems.  The    influence  of
gravitational lensing upon these models is considered and we find that
while the  emission from the central  continuum emitting region may be
significantly  enhanced,  lensing induced  magnification cannot easily
reconcile  the  models with  observations.   We conclude that  further
modeling,  including the effects of  any differential magnification is
required to explain the observed emission from \name.
\end{abstract}

\keywords{gravitational lensing -- infrared: galaxies --
quasars: individual(\name)}

\newpage

\section{Introduction}\label{introduction}
With bolometric  luminosities  exceeding $10^{12}\lsun$, Ultraluminous
Infrared Galaxies (ULIRGs) represent an extreme class of objects whose
spectra are dominated by  emission in the  far--infrared (for a review
of their  properties see~\cite{sa96}). The  source of their prodigious
output is thought  to arise  in  a thick, cool (T$\sim$100\K)  nuclear
dust structure which reprocesses emission  from an obscured  AGN--like
core,  a    massive   star--formation    region,  or    possibly  both
(\cite{ge98}).  The funneling  of  fuel  into   the nucleus and    the
resultant activity may be  triggered by a violent interaction  between
gas--rich galaxies (\cite{ba96,ta98});  this appears  to be the   case
observationally,  with  $\sim90\%$  of  ULIRGs displaying disturbed or
merging morphology (\cite{cl96}).

Recently identified in a survey of halo carbon stars, the z=3.87 Broad
Absorption  Line  (BAL) quasar, \name,   was found to  be positionally
coincident  with a   source  in    the  IRAS  Faint  Source    Catalog
(\cite{ir98}).    With  a flux  of 0.9Jy    at  100\micron, this ${\rm
m_r=15.2}$ object possesses  an inferred bolometric luminosity of $\rm
5\times10^{15}{\rm L_{\odot}}$ (${\rm \Omega_o =1}$ and ${\rm H_o = 50
km/s/Mpc}$ throughout), making  \name\  apparently  the most  luminous
object   currently known.   However,   ground--based images   taken in
$\sim0\scnd9$ seeing reveal   that the source  is slightly  elongated,
suggesting that \name\ consists of a pair of sub--components separated
by $\sim0\scnd3$,  consistent with  the merger--driven  hypothesis but
also with the action   of gravitational lensing.  In  several  extreme
ULIRGs the  morphology  of the   sub--components directly reveals  the
action of gravitational lensing, the magnification effect of which can
significantly enhance the intrinsic properties   of a system; this  is
the  case    in    both    H1413+117  (the   ``Cloverleaf     Quasar''
\cite{ma88,kn98})  and  the  hyperluminous   galaxy  IRAS  F10214+4724
(\cite{ro91,br95,ei96}), where  the  lensing-induced amplification  is
estimated to be  $\sim30-100$.   As \name\ is apparently  so extremely
luminous,  the possibility  that gravitational lensing  is influencing
the observed properties is highly likely.

Given  the   cool temperature   of the   dust region, photometric  and
spectroscopic observations at submillimeter wavelengths have proved to
be an   important  probe of ULIRG systems,    revealing details of the
physical properties and  processes underway in these energetic objects
(e.g.   \cite{iv98,iv98a,hu98}).  In   this  paper,   we  present  new
submillimeter photometry of \name, from which an estimate of the total
mass of dust, and  its temperature,are derived.  Combining  these data
with  previous observations,  we   compare the total  spectral  energy
distribution  to popular models of  ULIRG  galaxies, while taking into
account  the potential  effects  that  gravitational  lensing may   be
playing in distorting our view of this system.

\section{Observations}\label{observations}
The  observations were  conducted  with  the Submillimeter Common-User
Bolometer  Array (SCUBA,  \cite{ge95})   on  the James Clerk   Maxwell
Telescope~\footnote{The James Clerk Maxwell  Telescope is operated  by
The Joint   Astronomy  Centre on  behalf of  the  Particle Physics and
Astronomy Research Council   of  the United  Kingdom,  the Netherlands
Organisation for   Scientific   Research, and the   National  Research
Council  of Canada.}.   SCUBA  contains   a number  of  detectors  and
detector arrays  cooled to 0.1\K\ that  cover the  atmospheric windows
from 350\micron\ to 2000\micron.  For our photometric observations, we
operated the 91  element Short-wave (SW)  array at 450\micron, the  37
element Long-wave (LW) array at  850\micron, and the single photometry
pixel at 1350\micron, giving half-power  beam widths of 7.5, 14.7, and
$21\scmd$ respectively.  A    9-point jiggle pattern was employed   to
reduce the impact of  pointing errors. The  source was centered on the
central  pixel of  the arrays with  the  outer  pixels being used  for
subtraction of the sky variations.  Whilst jiggling, the secondary was
chopped at 7.8125\,Hz by $60\scmd$ in  azimuth. The pointing stability
was checked every hour  and regular skydips  were performed to measure
the atmospheric opacity.  The rms pointing errors were below $2\scmd$.

The  observations  were conducted over  3 nights  in April,  1998. The
first night had stable atmospheric zenith opacities at 450\micron\ and
850\micron, with $\tau$ being 1.47 and  0.29, but extremely high winds
which forced dome  closure. Firm  detections   were obtained at   both
wavelengths.  Sky conditions were reasonably  good on the second night
and observations  at both 850\micron/450\micron  ($\tau=0.58/3.7$) and
1350\micron\ were carried out. On the third night conditions were less
favorable (CSO $\tau$ around 0.1  and variable) and only 1350\micron\
observations   were done~\footnote{Observations on  days   2 \& 3 were
undertaken as part  of the  CANSERV  program operated by the  Herzberg
Institute for Astrophysics.}.

On the first and second nights  both CRL618 and IRC10216 were observed
as  calibrators,   while  only IRC10216 was    observed   on the third
night.  The latter is variable on  a timescale of   two years, and all
observations were referenced to CRL618, which has well-determined flux
densities with  SCUBA at 850\micron\  and 450\micron, but  nothing yet
published  for  SCUBA at 1350\micron. In  the  latter case, the CRL618
flux density published by Sandell  (1994) was used,  and resulted in a
sensible   value  for IRC10216. The  values    derived for IRC10216 at
450\micron/850\micron\ from CRL618  were also  well within the  quoted
errors.

The dedicated SCUBA  data  reduction software (SURF, \cite{je98})  was
used  to  reduce the  observations.  All  non-noisy  bolometers beyond
$40\scmd$  of    the central  pixel   were   used  to   compensate for
spatially-correlated  sky   emission   in the   850\micron/450\micron\
arrays. The     1350\micron\  pixel currently  has  no   provision for
subtracting sky variations  using the  other wavelength  pixels.   The
results of   these observations,  as  well as   variance-weighted mean
values, are listed in Table~\ref{table1}.

\section{Spectral Energy Distribution}\label{sed}
The  submillimeter  photometry   presented  in Table~\ref{table1}   is
combined  with previous data (Irwin  \etal 1998) in Figure~\ref{fig1},
which shows the spectral energy distribution  (SED) of \name.  The new
data reveals  that, in the submillimeter regime,  the SED  possesses a
slope of  $3.1\pm0.2\     {\rm  in\ \nu  F_\nu}$,    consistent   with
Rayleigh-Jeans   blackbody emission.     The  curves  superimposed  on
Figure~\ref{fig1} show  pure   blackbody   spectra of  a  source    of
temperature T=120\K (dot-dash  line) and  T=220\K (dotted line);  such
temperatures are  representative   of the  temperature range  of   the
regions  of the source  that emit  in  the submillimeter/far  infrared
region of   the SED.   These   values are slightly higher   than those
determined  for the  far--infrared  emission in   other ULIRG systems,
which are  typically T$\simlt$100\K\ (\cite{ea96}). The implied radius
of  such a blackbody is $\sim650$pc  (assuming the  emission region is
spherical),   a factor of  2  smaller  than the   cooler (80\K) region
responsible for the emission in IRAS F10214+4724~(\cite{do92}).

As outlined  in~\cite{hi83}, observations in  the submillimeter can be
used       to   determine    the    mass      in   dust     of  ULIRGs
(e.g.~\cite{cl92,do92,ea96}). Typically,   the dust emission  in these
systems appears to be optically thin,  as is inferred from their steep
SED slope  (the blackbody spectrum  is  modified by a  dust absorption
coefficient, ${\rm  \kappa_d\propto \nu^n}$,  where ${\rm  n=1-2}$; in
the  Rayleigh-Jeans regime,  this is  apparent as  a steepening of the
slope to at least 4 in $\rm  \nu F_\nu$).  The SED  of \name\ does not
show  such a steep slope, it  is instead consistent with emission from
an  optically  thick  blackbody. The   dust   mass obtained from   the
application of the optically thin treatment can, therefore, be treated
as a lower bound on the total dust mass in  the system.  Assuming that
the dust has a temperature of 220\K, the resulting  dust mass from the
three   submillimeter  data    points  presented    here  is     ${\rm
3.7\pm0.3\times10^9\msun}$  [assuming  a   constant   dust  absorption
coefficient of ${\rm \kappa_d}$  = 0.1 ${\rm  m^2/kg}$ (\cite{hu98})].
If     gravitational lensing  plays  a   role,    this value could  be
over--estimated by   a factor of $\sim30$,   although it does indicate
that \name\  possesses  copious amounts of  dust,  equivalent  to that
inferred   for IRAS  F10214+4724  (\cite{do92}), and   in other ULIRGs
(\cite{ea96}).

The complex  form  of   the SED  in  ULIRGs  suggests  that  a  single
temperature blackbody source   for  the FIR  flux   represents a gross
simplification     of   the   underlying   processes,    and   several
multi-component models  have  been  developed to  better  model  these
systems  (\cite{ro93,ef95,gre96,gra96}). These    models     generally
consider a powerful source of continuum radiation embedded in an thick
distribution   of   dust  possessing  grains  of   varying composition
(\cite{ro86}).   Radiative transfer techniques   are then  employed to
calculate the reprocessing  and  re-emission of the  central continuum
radiation.

Three such models are superimposed on the submillimeter-optical region
of the  SED in Figure~\ref{fig2}.  The dot-dashed curve represents the
``Embedded  Quasar    Model'' of~\cite{ro93}  (their  Model   C); this
possesses  a    power-law  continuum source    at   the center  of   a
spherically-symmetric dust distribution. As the dust obscures a direct
view  of  the illuminating  source,  all  observed emission  has  been
reprocessed. This model spectrum   is normalized to  the submillimeter
data points presented here.  Such  a  ``pure dust emission''  spectrum
provides  a  poor   fit  to \name\  and   other ultraluminous systems,
severely  underestimating the optical flux.    Spherical models with a
lower optical depth allow some continuum flux through the dust region,
but such models cannot account for the far infrared/submillimeter flux
(\cite{ro93}).

A natural extension  to such a  model  is to remove  the assumption of
spherical  symmetry and consider  instead an axisymmetric distribution
of absorbing dust,   representing a torus   about  the nuclear  region
(\cite{ro93,ef95,gre96}).  As  well as  scaling with total luminosity,
the emergent  spectrum  is also then  a  function of  orientation with
respect to the  observer; when  viewing a  system from  the equatorial
plane the dusty torus  obscures the view  of the continuum source  and
the SED is dominated by infrared dust emission.  When viewing from the
pole, however,     an observer can  look   directly  onto  the central
continuum source and the SED can possess both significant infrared and
optical components.  Such a  model was recently  applied to several of
extreme ultraluminous   systems and, considering  only orientation and
total luminosity, was found  to reproduce the gross characteristics of
the observed SEDs (\cite{gra96}).  The solid line in Figure~\ref{fig2}
of this  model, as viewed from  the pole, while  the dotted line is an
equatorial view.   Also  plotted  are the  rest   frame SEDs for   the
ultraluminous  Cloverleaf quasar,  H1413+117   (filled triangles), and
IRAS FSC  10214+4752 (crosses), both  normalized  to the submillimeter
observations of \name.  While  adequately describing these spectra, it
is apparent  that the models   of Granato  et al.  underestimates  the
infrared to optical flux by a factor of  $\simgt5$ even at the extreme
polar viewing orientation, and provide a poor fit of the SED of \name.
Extrapolating this  model  to the data presented  in  this system, the
implied  dust    mass is   $\sim3\times10^8\msun$,  but  again  likely
underestimates the  true mass of dust  in this system and is dependent
on the degree of gravitational lensing.

\section{The Influence of Gravitational Lensing}
The degree to which the flux  from a source is  enhanced by the action
of gravitational lensing is dependent on the scale-size of the source,
leading     to    pronounced    differential    magnification  effects
(\cite{sc92}); in IRAS F10214+4724 the optical  emission is thought to
be magnified by a factor  of $\sim100$, while the  flux from the  more
extended  far--infrared region undergoes  a magnification  of $\sim30$
(\cite{ei96}).  If  gravitational lensing  influences \name,  can such
effects  account  for the discrepancy  between  these  models  and the
observed   SED?  In   ``typical''   quasars,  the  optical-to-infrared
continuum  ratios are  seen  (in ${\rm  \nu  F_\nu}$)  to be $\sim  2$
(\cite{sa89,el94}). All this  emission arises in the central accretion
disk whose scale is  similar for both emission regimes;  gravitational
lensing would  uniformly magnify  such  a source, simply  scaling  its
contribution  to  the  total  spectrum.  Normalizing  the  (magnified)
quasar SED  to the observed  R-band magnitude it can  be seen that the
quasar continuum source, even if highly magnified  in this system, can
contribute  little to the submillimeter-infrared   SED.  It should  be
noted,   however,    that   if the   quasar    continuum  source does,
intrinsically, possess  significant  emission into the  far--infrared,
differential  magnification  effects may reconcile   the data with the
current models.

\section{Conclusions}\label{conclusions}
This  paper    has presented new   submillimeter observations   of the
ultraluminous BAL quasar \name; the  data are consistent with emission
from a  warm  (120\K-220\K), massive, optically thick  distribution of
dust which  is heated by a quasar  central continuum  source. A simple
model for the emission   region of ULIRGs,  consisting of  a AGN--like
continuum   embedded within    a spherical   distribution  fails, when
normalized to the submillimeter  data, to reproduce the  observed SED,
underestimating the the flux  in  the infrared  and, due to  a  highly
obscured view onto   the quasar  source,   predicts no  optical  flux.
Axisymmetric  models, where  the dust  is  distributed in a  torus, do
allow  relatively unobscured    views of the    quasar  core, allowing
unprocessed  optical flux to escape.  Even at the most extreme viewing
angles, however, current models still  underestimate both the infrared
and optical flux by several factors.

Further modeling of the geometry and physics of the emission region in
\name, coupled with the effects of possible differential magnification
due to gravitational lensing,  is  therefore required.  Such  modeling
will provide a more accurate determination  of the dust mass in \name,
which, when coupled with  spectroscopic observations of warm molecular
gas tracers  such as CO, will  shed  light on  the interaction between
star--formation, warm dust and the AGN core in explaining the observed
properties of this phenomenal system.

It  is interesting to note  that pronounced infrared emission has been
observed in a  number of Seyfert Is  and IIs  in the nearby  universe.
(\cite{bo97}).  Similar to  the  high redshift  ULIRGs, the source  of
this emission is thought to be a warm, dusty torus, with viewing-angle
dependent obscuration of the  continuum source and broad emission line
region accounting for the difference between  the Seyfert classes. The
similarity  between this model and that  proposed for more distant and
powerful  ULIRGs  (\cite{ba95}),  especially considering that  several
Seyfert I systems exhibit absorption  features indicating the presence
of high-velocity outflows [e.g. Markarian 231 (\cite{fo95})], suggests
that, rather than being  unique objects displaying unusual properties,
systems such as \name\ and H1413+117 appear to be simply more luminous
members of the ULIRG/AGN family.

\section{Acknowledgments}\label{ack}
We thank Henry Matthews of the JCMT for  the acquisition and reduction
of   the CANSERV sub--millimeter data, and    Paul Feldman and Russell
Redman for details of the  CANSERV application.  The anonymous referee
and Zdenka Kuncic are thanked for useful comments.

\newpage

\begin{figure}
\centerline{
}
\caption[]{The SED for \name. The open circles present the data
presented in this paper. The filled circles are detailed
in~\cite{ir98}, with the arrow indicating an upper-limit.  The two
curves are the SED for pure blackbody emitters, the solid curve for a
system of temperature T=220K, and the dot-dashed curve for T=120K. The
curves have been normalized to the submillimeter observations
presented in this Letter.}
\label{fig1}
\end{figure}

\newpage

\begin{figure}
\centerline{
}
\caption[]{The submillimeter to  optical SED of \name.  The dot-dashed
curve represents  the  SED  for  a  quasar  source embedded   within a
spherical distribution   dust  (\cite{ef95}),  while the solid   curve
represents a face-on  view of a quasar at  the center of a dusty torus
(\cite{gra96}).  The dotted line  represents this latter  model viewed
in  the equatorial  plane.  The  curves have  been  normalized to  the
submillimeter  observations presented in   this Letter.  Also included
are  the  emission-frame SEDs for H1413+117   (triangles) and IRAS FSC
10214+4724 (crosses),  normalized to the  submillimeter  SED of \name.
As can  be seen, none of  the  models adequately  represent the SED of
\name, while those of Granato describe SEDs of the other ultraluminous
systems.  }
\label{fig2}
\end{figure}

\clearpage
 
\begin{deluxetable}{cccc}
\footnotesize \tablecaption{ Submillimeter photometry of APM
08279+5255. The bottom line presents the variance-weighted means of
the observations.
\label{table1}}
\tablewidth{245pt}
\tablehead{
\colhead{Date} & \colhead{$\thirteen$} &  \colhead{$\eight$} &
\colhead{$\four$}
}
\startdata
4$^{\rm th}$April  & ----        & $75\pm4$mJy & $203\pm51$mJy  \nl
18$^{\rm th}$April & $26\pm4$mJy & $74\pm9$mJy & $260\pm130$mJy \nl
19$^{\rm th}$April & $23\pm3$mJy & -----       & -----          \nl
                   & $24\pm2$mJy & $75\pm4$mJy & $211\pm47$mJy  \nl
\enddata
\end{deluxetable}

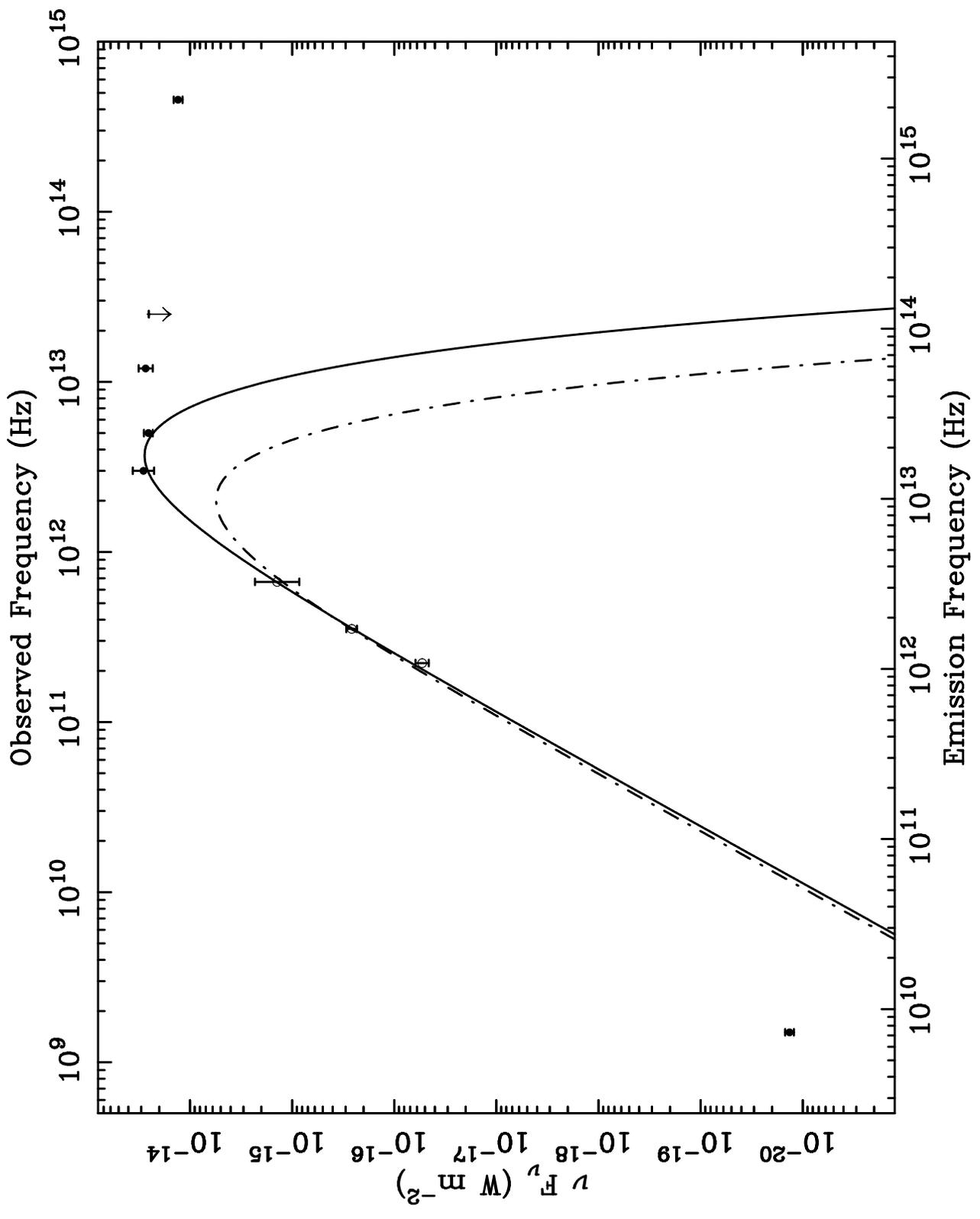
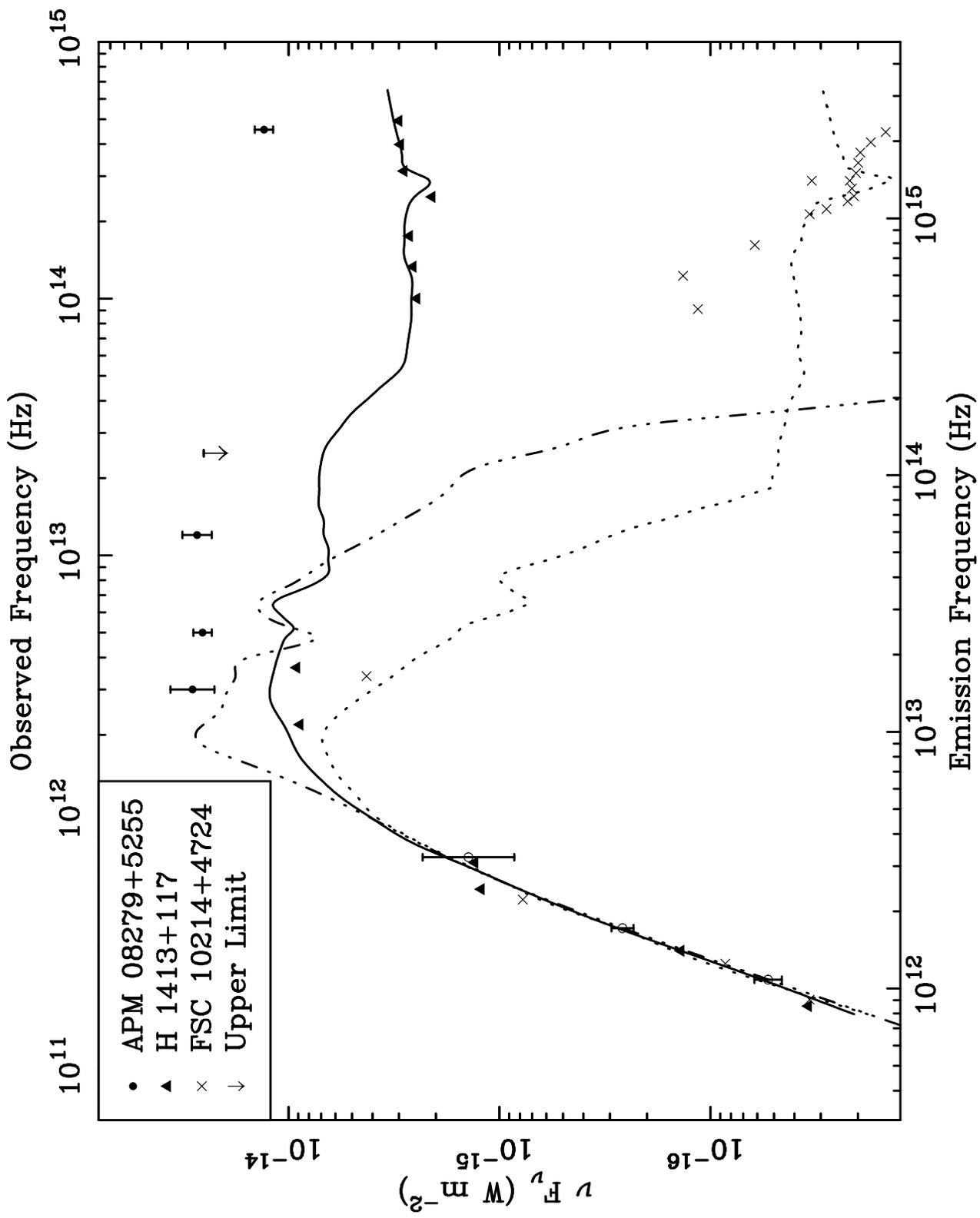
\end{document}